# Positron lifetime calculation of the elements of the periodic table

**J M Campillo Robles[1], E Ogando[2] and F Plazaola[2]**

[1] *Oinarrizko Zientziak Saila, Goi Eskola Politeknikoa, Mondragon Unibertsitatea, 20500 Arrasate, Basque Country.*

[2] *Elektrizitate eta Elektronika Saila, Zientzi Fakultatea, 644 P.K. UPV-EHU, 48080 Bilbo, Spain.*

E-mail: jmcampillo@eps.mondragon.edu (J M Campillo Robles), edu.ogando@ehu.es (E Ogando) and fernando@we.lc.ehu.es (F Plazaola).

**Abstract.** Theoretical positron lifetime values have been calculated systematically for most of the elements of the Periodic Table. Self-consistent and non-self-consistent schemes have been used for the calculation of the electronic structure in the solid, as well as different parameterizations for the positron enhancement factor and correlation energy. The results obtained have been studied and compared with experimental data, confirming the theoretical trends. As it is known, positron lifetimes in bulk show a periodic behaviour with atomic number. These calculations also confirm that monovacancy lifetimes follow the same behaviour. From this fact a strong relation between the atomic volume and the positron lifetime has been set. The effects of enhancement factors used in calculations have been commented. Finally, we have analysed the effects that f and d electrons cause in positron lifetimes.

## 1. Introduction

The classification of the elements has been one of the major achievements in the history of Science. This classification was originally derived from empirical experimental results, because the concept of atomic number was unknown to Mendeleev [1]. Since then, the resulting periodic order has been most strikingly reflected in a quantitative manner by most of the physical properties of the elements. A proof of this fact is that about 700 forms of the Periodic Table have been proposed (classified into 146 different types or subtypes) [2, 3].

Positron-Annihilation Spectroscopy (PAS) is a powerful and versatile tool for the study of the microscopic structure of materials [4-7]. Using PAS, detailed experimental information about electronic and atomic structure from the region of the material sampled by the positrons is obtained. PAS measurements for material characterization generally use three techniques: positron lifetime spectroscopy, Doppler broadening analysis and angular correlation measurements. Positron lifetime measurements give information about electron density at the annihilation place. On the other hand, Doppler broadening and angular correlation measurements give information about electron momentum distribution. So, the electronic configuration of the material studied is reflected in the positron annihilation parameters. For example, in Doppler broadening experiments, the high-momentum part is used to distinguish different non-adjacent elements in the Periodic Table [8, 9].

As the annihilation properties of the positron are related to the electronic structure, they also show a periodic behaviour. In 1963, Rodda and Stewart [10] studied the behaviour of the experimental positron lifetime of rare-earth metals and compared it with the radius of the sphere, whose volume is equal to the volume per conduction electron, $r'_s$ (correcting the value by excluding the volume of the ion). However, MacKenzie et al. set a stronger relation in 1975 [11]. They collected experimental bulk lifetimes of many elements and reported their "systematic dependence on atomic number". Also that year, Brandt et al. [12] calculated the lifetime of some elements. They explained that the periodic behaviour of positron lifetime "is linked *prima facie* to the virtual excitation of coupled valence-electron-plasma and single-particle modes in the collective response of the metal electrons to screen the positron charge". Later, in 1976, Welch and Lynn studied the variations of the experimental mean lifetime versus the atomic number, and stated that it "is strikingly similar to that of the atomic volume" [13]. In 1991 Puska found that the trends observed in bulk lifetimes along the 3d, 4d and 5d rows of the Periodic Table are very similar to the behaviour of the Wigner-Seitz radii [14]. However, this periodicity is not only reflected in the positron lifetime. Doppler broadening experiments [9, 15] and positron affinity of elemental metals [16] also show periodic behaviour.

In this work, a systematic Density Functional Theory (DFT) calculation of positron lifetimes has been performed for bulk and monovacancies of most of the elements of the Periodic Table. The main factors influencing bulk and vacancy positron lifetimes for elemental solids have been well understood for more than 30 years. So, our main aim is to show the periodic trends appearing in bulk and monovacancy positron lifetimes. The effort made to calculate and compile systematically the annihilation parameters is important to go deeply into the study of the calculation methods, improving the theoretical background required for a good interpretation of the experimental data. The organization of this paper is as follows: the computational method is explained in section 2; section 3 contains the results of the calculations, the correlation between lifetimes, and the periodic properties of the elements, information about the enhancement factors used in the calculations and the analysis of the effects of f and d electrons; and finally, the conclusions of the work are presented in section 4.

## 2. Computational method

The calculation of positron properties in solids can be traced back to the late sixties and early seventies (see for instance [17-20]) and since then numerical simulations have become a well developed technique (see the reviews [4, 6] and the recently published paper [21]). Positron states must be calculated self-consistently within the two-component DFT for positron and electron densities. However, the conventional way to treat positron states in solids simplifies the two-component DFT. Within the conventional scheme, an unperturbed electronic ground state for the system is constructed. Then, the positron distribution is calculated by assuming the electron density remains rigid, and by accounting for the electron-positron correlation in terms of a correlation (screening) potential dependent on the electron density. In the case of delocalized positron states, the positron density is vanishingly small at every point of the lattice, and it does not influence the electronic structure. As a result, for bulk positron states the conventional scheme runs very well. In the case of a positron localized at a lattice defect, the situation is more complicated because the positron attracts electrons, and the average electron density increases near the defect (positron). However, in most applications for positron states at defects, the conventional scheme works very well too. Indeed, the two-component DFT calculations performed by Nieminen et al. [22] and Boronski and Nieminen [23] support the use of the conventional scheme, since the annihilation rates are very close to those obtained with the conventional scheme. This similarity of results between the conventional scheme and the two-component DFT calculation is due to the fact that the larger short-range enhancement compensates the smaller electron density at the positron.

Therefore, in the present work we have used the conventional way of calculation. Firstly, we have solved the electron density of the perfect or defected solid, then we have calculated the positron wave function, and finally, we have determined the positron annihilation rate. We have used a supercell method to compute the electronic densities following (a) the atomic superposition approximation (AT-SUP) developed by Puska and Nieminen [24] and (b) the tight binding version of the Linear Muffin-Tin Orbital method within the Atomic-Spheres Approximation (LMTO-ASA) [25, 26].

(a) AT-SUP method

The AT-SUP approximation of Puska and Nieminen is a simple method that makes use of non-self-consistent unrelaxed electronic densities. It gives satisfactory values of positron lifetimes in metals and semiconductors [24, 27, 28, 29]. The good agreement between the experimental and theoretical lifetimes is mainly due to the fact that the positron annihilation rate is obtained as an integral over the product of positron and electron densities. The positron density relaxes following the electron charge transfer, keeping the value of the positron-electron overlap integral constant. For this reason, positron lifetime calculations are not too sensitive to self-consistency.

In the AT-SUP approximation the electron density $n_-(\mathbf{r})$ of the solid is constructed by superimposing individual atomic charge densities:

$$n_-(\mathbf{r}) = \sum_i n_-^{at}\left(\left|\mathbf{r} - \mathbf{R}_i\right|\right) \qquad (1)$$

where $n_-^{at}$ is the free-atom electron density and $\mathbf{R}_i$ runs over the occupied atomic sites. For the crystalline Coulomb potential $V_c(\mathbf{r})$ the same procedure has been used:

$$V_c(\mathbf{r}) = \sum_i V_{at}\left(\left|\mathbf{r} - \mathbf{R}_i\right|\right) \qquad (2)$$

where $V_{at}$ is the atomic Coulomb potential due to the electron density and the nucleus. Densities and potentials of the atomic ground-state electronic configuration are obtained self-consistently within the DFT.

The potential felt by the positron in the solid, $V_+(\mathbf{r})$, is obtained by adding to the Coulomb potential, $V_c(\mathbf{r})$, the positron-electron correlation energy, $V_{corr}(n_-(\mathbf{r}))$:

$$V_+(\mathbf{r}) = V_c(\mathbf{r}) + V_{corr}(n_-(\mathbf{r})) \tag{3}$$

where $n_-(r)$ is the electron density. The space is discretized in a three-dimensional mesh that forms an orthorhombic Bravais lattice, where the potential is projected. The discretized Schrödinger equation is solved iteratively at the mesh points by using a numerical relaxation method [30] to obtain the positron wave-function and its energy eigenvalue. Depending on the structure of the element, the density of the cubic mesh varies between 1 and 3 points per atomic unit in each direction. We have checked in some elements that this difference of density of the mesh does not affect the lifetimes.

(b) LMTO-ASA method

LMTO-ASA is a method that makes use of self-consistent electronic densities. It gives satisfactory values of positron lifetimes in metals and semiconductors [14, 28, 29]. In LMTO-ASA calculations the electron density and Coulomb potential are determined self-consistently within spheres centred around nuclei and interstitial sites (when the atomic packing is not dense) of the structure. The spheres fill the whole lattice space and the atomic ones have equal radii. The potential and the charge densities are assumed spherically symmetric inside each sphere. The potential felt by a positron is constructed according to equation (3), and the positron state is solved by using the same methods as used for electron states in the LMTO-ASA. From now on, all the references made to this method will be labelled as LMTO.

The f electrons are a strongly correlated system. However, we have treated the 4f electrons of the lanthanides and the 5f electrons of the actinides as band electrons, and the 4f electrons of the actinides as core like states.

Once we have calculated the electron and positron densities, the positron annihilation rate, the inverse of the positron lifetime, is obtained from the overlap of positron and electron densities as:

$$\lambda = \pi r_o^2 c \int \mathrm{d}\mathbf{r} n_+(\mathbf{r}) n_-(\mathbf{r}) \gamma(\mathbf{r}) \tag{4}$$

where $r_o$ is the classical electron radius, $c$ is the speed of light in the vacuum, $n_+(\mathbf{r})$ is the positron density and $\gamma(\mathbf{r})$ is the so-called enhancement factor. $V_{corr}(\mathbf{r})$ and $\gamma(\mathbf{r})$ have been taken into account by using two different schemes:

1.- within the Local Density Approximation. For the correlation energy the interpolation formula by Boroński and Nieminen [23] based on the results by Arponen and Pajanne [31] is used; and for the enhancement factor the widely used form [23] based on Lantto´s [32] hypernetted chain approximation calculations:

$$\gamma_{BN}(r_s) = 1 + 1.23 r_s + 0.8295 r_s^{3/2} - 1.26 r_s^2 + 0.3286 r_s^{5/2} + \frac{1}{6}\left(1 - \frac{1}{\varepsilon_\infty}\right) r_s^3, \qquad (5)$$

where $\varepsilon_\infty$ is the high-frequency dielectric constant of the material and $r_s$ is the radius of a sphere whose volume is equal to the volume per conduction electron. This last parameter is related to the electron density, $n_-(\mathbf{r})$, by:

$$r_s = \left(\frac{3}{4\pi n_-}\right)^{1/3}. \qquad (6)$$

Results obtained with this scheme are labelled with BN.

2.- within the Generalized Gradient Approximation (GGA). The correlation energy and the enhancement factor due to Barbiellini et al. [28, 33] are used, both based on the results by Arponen and Pajanne [31]. In this scheme the enhancement factor is given by:

$$\gamma_{GGA} = 1 + (\gamma_{LDA} - 1) e^{-\alpha\varepsilon}, \qquad (7)$$

where $\gamma_{LDA}$ is:

$$\gamma_{LDA}(r_s) = 1 + 1.23 r_s - 0.0742 r_s^2 + \frac{1}{6} r_s^3, \qquad (8)$$

and $\alpha$ is an adjustable parameter and $\varepsilon$ is obtained from this expression:

$$\varepsilon = \frac{|\nabla n_-|^2}{(n_- q_{TF})^2} = \frac{|\nabla \ln n_-|^2}{q_{TF}^{\ 2}}, \qquad (9)$$

with $(q_{TF})^{-1}$ the local Thomas-Fermi screening length. $\varepsilon$ is a parameter proportional to the lowest–order gradient correction to the correlation hole density in the Local Density Approximation. The results obtained with these two schemes will be labelled LDA and GGA.

The $\alpha$ parameter is determined so that the calculated and experimental lifetimes agree as well as possible for a large number of different types of solids. Barbiellini et al. [28, 33] found that $\alpha$ = 0.22 value gives lifetimes in good agreement with experiments for different types of electronic environments, including simple metals (Na and K of 1st group), transition metals (Fe, 8th group; Ni, 10th group; Cu, 11th group; Al, 13th group), group-IV semiconductors (Si and Ge) and III-V and II-V compound semiconductors (GaAs, InP and CdTe). $\alpha = 0$ gives the Local Density Approximation limit of this enhancement factor, that is to say $\gamma_{LDA}$. As it was pointed before [28, 34], the lifetimes calculated by using $\gamma_{LDA}$ are always much shorter than those calculated with $\gamma_{BN}$ and $\gamma_{GGA}$, and the experimental ones.

The positron lifetime calculations have been performed for most of the elements of the Periodic Table. The unit cell of the crystalline structure has been used as supercell for bulk calculations. In vacancy calculations one atom is removed from the supercell to produce a vacancy. If the supercell is large enough, the vacancy does not interact with its periodic image and the system describes quite well an isolated vacancy. However, in practice, the supercell size cannot be made arbitrarily large. The size of the supercells in AT-SUP method has been increased till convergence. The maximum number of atoms per supercell used to reach convergence has been: 511 atoms (orthorhonbic structure), 511 atoms (diamond structure), 499 atoms (tetragonal structure), 463 atoms (cubic structure), 255 atoms (FCC structure), 249 atoms (hexagonal structure), 383 atoms (rhombohedric structure) and 127 atoms (BCC structure). Self-consistent calculations within the LMTO method for monovacancies are much more computationally demanding than the ones performed with the AT-SUP method, particularly for large supercells. Moreover, for large supercells, calculations performed employing the Γ point for the positron density or the lowest lying band in the Brillouin zone give identical values for the monovacancy positron lifetime [29]. Therefore, for the calculations within the LMTO method for monovacancies we have integrated over the lowest positron band in the Brillouin zone, because it gives faster convergence in the supercell approach [29]. The maximum number of atoms used within the LMTO has been: 127 atoms (hexagonal structure), 124 atoms (BCC structure), 107 atoms (rhombohedric structure), 63 atoms (FCC structure), 53 atoms (tetragonal structure), 53 atoms (diamond structure), 31 atoms (orthorhonbic structure) and 28 atoms (cubic structure).

For the monovacancy supercells no relaxation in the atomic positions have been performed, this means that the ions neighbouring the vacancy are not allowed to relax from their ideal lattice positions. It is known that an accurate calculation of the monovacancy lifetime needs atomic relaxation. In insulators and semiconductors the atomic relaxation may be important, and can change with the charge state of the defect and with the localization of the positron [35, 36], but it is not large in metals [37-39]. However, the study of the effects of these relaxations goes beyond the aims of this work.

Some elements get a different crystal structure for different conditions of pressure and temperature. When an element has more than one possible structure, we have chosen the most common one in normal conditions. The rare gases are not solid in normal conditions, so we have studied the solid state at very low temperature. The crystal structure and the lattice parameters used in the calculations are shown in table 1, and have been taken from experimental data [40-42].

In BN calculations, the semiempirical correction based on high-frequency dielectric constant of equation 5 [43] is used for the elements of table 2, which shows experimental values of dielectric constants [41, 42, 44]. For several insulators (As, Cl, Br and I) we have not found any value for their dielectric constants in the literature. So, these insulators and rare gases have been treated as metals, using $\varepsilon_\infty = \infty$. In the rare gases a special treatment is needed; however, it goes beyond the aim of this work. In LDA and GGA frameworks, this correction is not necessary.

## 3. Results and discussion

### *3.1. Bulk and monovacancy lifetimes*

The positron lifetime values calculated for the bulk state are given in table 3 and table 4. The values of table 3 correspond to calculations made within the AT-SUP method with BN and GGA approximations. The results obtained within the LMTO method with BN and GGA approximations are given in table 4.

On the other hand, the results of monovacancy lifetime calculations (BN and GGA) are shown in tables 5 and 6 for AT-SUP and LMTO methods, respectively. Table 6 do not present monovacancy lifetime of actinides due to convergence problems with the LMTO code. The bulk and monovacancy lifetime results are in agreement with previously reported values (Barbiellini et al. [28, 33]).

Table 7 shows the AT-SUP results of bulk and monovacancy lifetimes obtained within the LDA framework. As mentioned before, these values are shorter than BN and GGA ones, but show the same trend.

Finally, experimental positron lifetime values in bulk and monovacancy states (see [45] and the references therein) are given in table 8 for comparison with theoretical ones. Even though the first positron lifetime measurements were made more than 50 years ago [46-49], nowadays the experimental data does not reach to all elements of the Periodic Table. Moreover, there is much more data for bulk than for monovacancy lifetimes. In some elements, there is no experimental data, but in others the experimental data and the scattering among them is large. Therefore, the selection of the measurements is a difficult affair, and we have fixed some conditions to select data with a minimum of quality and coherence. The chosen conditions might not be the best ones; however, a selection has to be made. First of all, we have considered data from 1975 up to now. We have chosen this requirement because the POSITRONFIT program was developed around 1972 [50], and improved in 1974 [51], becoming a common, or even standard, tool for the positron community to analyze experimental spectra. Furthermore, we have chosen a maximum of 320 ps for the full width at half maximum of the resolution function. Finally, we have taken as the limit value for the error of the measurement ±5 ps in bulk lifetimes, and ±10 ps in monovacancy lifetimes. In the case the literature gives different lifetime values following the previous requirements, the average value has been calculated. It is expected that the systematic errors from various experiments would be cancelled. In order to fill the extremes of the Periodic Table, we have taken into account two experimental works in alkalines (Li, Na, K, Rb, Cs) [52] and ideal gases (Ar, Xe) [53] that do not fit the previous requirements. However, these works have been used in previous reviews.

### *3.2. Positron lifetime among periodic properties*

In order to present the expected periodic behaviour of the positron lifetime, figure 1 shows calculated bulk (circles) and monovacancy (squares) positron lifetimes versus atomic number. Plotted lifetimes have been calculated within the AT-SUP method using BN approximation for the enhancement factor and correlation energy. Experimental values of table 8 have not been plotted, but follow the same theoretical trends.

The atomic volume of the elements [54] has been plotted against atomic number in figure 1, too. The atomic volume (defined as the product of the atomic weight and the specific volume of an element at normal conditions) is a good magnitude to measure the size of one single atom in its own structure and is defined for all the elements in the same way. This is one of the main reasons why the atomic volume has been chosen, even though different magnitudes such as metallic radius, ionic radius, covalent radius ... have been defined for the quantification of the atomic size. However, from these last magnitudes it is not possible to obtain an accurate value for the volume. Besides, most of radii types are defined only for some kind of elements and not for all. The atomic volume, as other properties of the elements, has a strong relation with the arrangement of the electrons in atomic shells [55-59]. For this reason, the atomic volume is a periodic function of the atomic number, as it was formulated first by Lothar Meyer in 1870 (with reference to atomic weight, not to atomic number) [60].

The similarity between atomic volume curve and the two positron lifetime curves on figure 1 is very big. Different factors affect positron lifetimes, like many-body enhancements, the region occupied by the positron (which in the bulk is less than the atomic volume), and the electrons available for annihilation in that region. However, the three graphics show the same periodic behaviour, reproducing also many little details. Although the lifetime has been compared with other periodic properties like $r_s$ parameter, Wigner-Seitz radii,... the relation with the atomic volume seems to be more fundamental. This work confirms previous statements [13], but also proves that the monovacancy lifetimes exhibit the same periodic behaviour. Despite the localization of the wave function, the positron lifetime in bulk (delocalized state) and at a monovacancy (localized state) are still related to a single atom's volume, and this is independent of the methods of calculation used in this work. When an atom is removed from perfect crystal structure, the remaining volume is mainly related to that atom. But removing more than one atom the remaining volume is more structure dependent [34].

The lifetimes of some elements (As, Br, Kr, I and Xe) do not follow the trends of atomic volume (see figure 1). This special behaviour will be analyzed in section 3.3.

The periodic behaviour of the positron lifetime found for AT-SUP method within BN approximation is also found using GGA and LDA frameworks, as well as in the calculations performed with the LMTO code by using BN and GGA approximations.

*3.3. Enhancement factors*

The enhancement factor is of crucial importance in positron lifetime calculations [61, 62, 63 and references therein]. For this reason, it is necessary to study the behaviour of enhancement factors used in these calculations.

As it has been pointed before, in figure 1 the lifetimes of some elements (As, Br, Kr, I and Xe) do not follow the periodic trends as it could be predicted from the atomic volume. In the case of insulators, a model based in atomic polarizabilities, estimated from the Clausius-Mossotti relation, has been used too [43], where the dielectric constant of the solid is also needed. For the rare gases a special framework is needed. However, in these calculations, some insulators (As, Cl, Br and I) and all the rare gases (Ne, Ar, Kr and Xe) have been considered as metals ($\varepsilon_\infty = \infty$). So the real lifetimes of BN approximation for these eight elements are really longer and closer to GGA values than the calculated ones (see tables from 3 to 6).

Boroński-Nieminen enhancement, $\gamma_{BN}$ (eq. 5), is based on the many-body calculations performed by Lantto [32]. Stachowiak and Boroński reported that the calculations of Lantto start from a physically oversimplified trial function [64]. Fraser in her Ph.D. thesis failed to reproduce Lantto's results using a Quantum Monte Carlo approach [65]. However, Boroński reported that the lifetimes calculated based on Fraser's results are quite inaccurate, becoming even unreasonable for $r_s > 4$ [66].

Boroński-Nieminen parameterization, used in this work, agrees very well the positron enhancements calculated by Arponen and Pajanne [67], Gondzik and Stachowiak [68] and Rubaszek and Stachowiak [69] for $r_s \leq 8$ (see Fraser´s thesis [65], page 143). Indeed, Stachowiak and Boroński [64] pointed out that $\gamma_{BN}$ is the best formula to fit the experimental lifetimes in metals. In bulk metals, $r_s$ usually runs from 2 to 6 (in Cs it gets the maximum value, 5'6). However, the enhancement factor of Boroński-Nieminen approach has two important problems at low densities (see figure 1 in [28]):

a) The scaled proton limit rule [70] is violated for $r_s \geq 9$. This is the upper bound for all the enhancement factors.

b) For $r_s$ greater than 6 the lifetimes obtained with $\gamma_{BN}$ do not increase monotonically with $r_s$. For this reason, the lifetimes can not reach the 500 ps limit.

These two problems appear when the electronic density is low (semiconductors, insulators, rare gases, vacancies, voids,…). In the case of semiconductors and insulators, the problem has been tackled using semiempirical corrections introduced by Puska et al. [43]. Using this correction, calculated lifetimes in the BN approximation fit well the experimental ones, even when the densities are low.

In the case of $\gamma_{LDA}$ (eq. 8), the enhancement factor has the same form as that used by Stachowiak and Lach [71]. $\gamma_{LDA}$ has been obtained fitting the Arponen-Pajanne data points only up to $r_s = 5$ [28]. In Arponen-Pajanne data, the Friedel sum rule is violated for $r_s = 6$ and 8, and the scaled proton limit value is crossed at $r_s = 8$ [31]. As $\gamma_{GGA}$ (eq. 7) is obtained from $\gamma_{LDA}$, the LDA and GGA enhancements are both not very reliable for low electron densities, like BN enhancement. Calculations made in systems of low electron densities using GGA approximation with the universal value $\alpha$ = 0'22 do not give reasonable lifetimes [63, 72]. For example, in order to get lifetime values near the experimental ones for $C_{60}$, it is necessary to fit the $\alpha$ parameter to a "suitable" value [72].

As a result of all of these problems in the enhancements, we have to be very careful with all of the calculated lifetimes near or longer than 400 ps. So, more theoretical work is needed for low density systems.

*3.4. f and d electrons*

Figure 2 represents the behaviour of positron lifetimes for bulk (circles) and monovacancy (squares) in elements from $_{57}$La to $_{80}$Hg of the 6th row of the Periodic Table. As f and d shells get filled between $_{57}$La and $_{80}$Hg, positron lifetimes of these elements can be used for studying the effects of f and d electrons in positron annihilation properties. The represented lifetimes are those obtained within the LMTO method using BN (empty symbols) and GGA (full symbols) approximations (see table 4 and 6). Experimental lifetimes from table 8 have been plotted as a reference. As it is known, GGA calculation method uses the gradient of electronic density. So, for an accurate calculation, it is necessary to use a self-consistent electronic density, enabling charge-transfer in the system. Therefore, in order to make a better comparison between BN and GGA calculation methods, we have represented LMTO results.

First of all, we must remark that figure 2 shows the same general behaviour for bulk and for monovacancy lifetimes. As electrons fill the d shell between $_{71}$Lu and $_{80}$Hg, the bulk and monovacancy lifetimes show the same parabolic behaviour. As d electrons start filling the shell, the lifetime reduces considerably, and, after reaching a minimum near the half-filled shell ($d^6$), it rises up again. This trend is explained simply with the behaviour of the atomic volume (see figure 1). This general dependence of positron lifetime with the outermost d electrons in bulk and monovacancies is independent of the row of the Periodic Table. The 4th row, from $_{21}$Sc to $_{30}$Zn elements, and the 5th row, from $_{39}$Y to $_{48}$Cd elements, show the same behaviour.

On the other hand, in lanthanides (from $_{57}$La to $_{70}$Yb) the positron lifetime remains nearly constant as 4f shell fills up ($\tau_{bulk} \approx$ 200 ps and $\tau_{vacancy} \approx$ 315 ps). The increasing number of the f inner electrons is responsible for the magnetic properties of lanthanides, and the outermost s-d electrons determine the bonding and other electronic properties [73-75]. So, the f inner electrons can not cause appreciable changes in positron lifetimes, since the positron wave-function is mainly located in the interstitial space. However, the lifetimes of $_{63}$Eu and $_{70}$Yb are larger than the other lanthanides, because they have a half-filled ($_{63}$Eu) or completely filled 4f shell ($_{70}$Yb). For this reason, Eu and Yb get a more closed electronic structure and show a particular behaviour in several properties (atomic volume, electronegativity, melting point, ionization potentials,...). It has to be remarked that in the case of actinides (from $_{90}$Th to $_{97}$Bk), the positron lifetime does not remain constant (see table 4). Indeed, it follows a parabolic behaviour (see figure 1) similar to the one found in d shells. This behaviour is in agreement with previous statements [76-78], which indicate that opposite to 4f electrons, 5f electrons are relatively delocalized and can contribute to the bonding. This special electronic structure makes positron lifetime (and other physical properties) behaviour more complex.

However, there are some especial features superposed to these general trends. In bulk, for the first lanthanide elements (from $_{57}$La to $_{62}$Sm) GGA lifetimes are similar to BN ones. But, between $_{64}$Gd and $_{69}$Tm, GGA lifetime is a little bit larger than BN one, about 4 ps (see table 4). There are only three experimental bulk lifetimes from the literature. These experimental lifetimes are from different researcher groups and not from very recent measurements. So, it would be interesting to get new experimental data in the lanthanides.

In monovacancies, lifetimes of lanthanides have a particular behaviour (see figure 2). For most of the elements of the Periodic Table, GGA lifetimes are usually larger than BN ones (see table 6 ). Opposite to this general trend, the GGA lifetimes of lanthanides are shorter than BN ones, about 3-5 ps (see table 6). As it is known, the positron is much localized in a monovacancy and the probability of annihilating with inner electrons is much lower than in bulk. Taking into account that f electrons are inner electrons, they have not an appreciable effect in these lifetimes. So, the values of these lifetimes are due to the external electronic configuration, similar to the outer electronic configuration of $_{57}$La. As in bulk case, $_{63}$Eu and $_{70}$Yb are outside this trend, showing a special feature. It could be expected them to follow the same trend as the other lanthanides. However, they show the opposite, GGA lifetimes are longer than BN ones.

A lot of work has been made to understand the behaviour of d electrons [79, 34]. In the bulk case (see figure 2), for La, Ce and Lu the BN lifetimes are longer than the GGA ones. However, as electrons start to fill the d shell, the difference gets smaller. And from $d^3$ ($_{73}$Ta) to $d^{10}$ ($_{80}$Hg) GGA lifetime is longer than BN one, increasing the difference as d orbitals are being occupied. The GGA correction to the Local Density Approximation is roughly proportional to the number of outermost d-electrons in the atom [34]. The four experimental lifetimes found in the literature fit very well GGA values in this region. Local Density Approximation calculations made with different parameterizations, BN (table 3) and LDA (table 7), show that the positron lifetimes for bulk transition metals are systematically too short in comparison with the experimental values, and GGA fits better experimental values.

In monovacancies the trend is similar to that of the bulk. BN lifetime is longer than GGA one for the first elements, from Lu ($d^1$) to Ta ($d^3$). At W ($d^4$) and Re ($d^5$) BN and GGA lifetimes are the same. And finally, from Os ($d^6$) to Hg ($d^{10}$) GGA lifetimes are longer than BN ones. So, the trend is the same, but in monovacancies more d electrons are needed for GGA lifetimes to get longer than BN ones. In monovacancies, the calculated lifetimes and the experimental ones show a larger difference than in the bulk case.

In lifetimes calculated within AT-SUP method, BN and GGA values follow the very same general trends. However, the differences between GGA and BN lifetimes are much greater, due to the lack of self-consistency of the electronic densities used in calculations.

## 4. Conclusions

The systematic calculations performed in this work set a theoretical support for understanding and interpreting different positron lifetime experiments. It has to be remarked that there are many firstly calculated elements among these lifetimes.

As a result of this positron lifetime calculations, a well-known trend for the bulk lifetimes has been systematically proved again, and the same trend has been established for the calculated monovacancy lifetimes too. In both cases, the correlation with atomic volume of the elements is direct, and it is demonstrated that the trends of atomic volumes can be extrapolated to positron lifetime values. However, a direct quantitative extrapolation on the absolute values can not be done in the whole Periodic Table due to the fact that positron lifetimes reach saturation at 500 ps. This fact deforms the trends compared to the atomic volume. So, it is concluded that the positron lifetime of bulk and vacancy is a periodic property of the elements.

**Acknowledgment**

The authors acknowledge N. de Diego and J. del Rio for some useful ideas for the work. We also want to thank J. M. Igartua for interesting discussion about structural data of some elements, and X. Artetxe for technical assistance.

**References:**

[1] Mendeleev D 1869 Zeitschrift für Chemie **12** 405.

[2] van Spronsen J W 1969 *The Periodic System of the Chemical Elements* (Elsevier, Amsterdam).

[3] Mazurs E G 1974 *Graphic Representation of the Periodic System during One Hundred Years* (University of Alabama Press, Alabama).

[4] West R 1973 Adv. Phys. **22** 263.

[5] 1979 *Positrons in Solids*, ed.: P. Hautojärvi (Springer Verlag, Heidelberg).

[6] Puska M J and Nieminen R M 1994 Rev. Mod. Phys. **66** 841.

[7] 1995 *Positron Spectroscopy of Solids*, ed.: A. Dupasquier and A. P. Mills, Jr. (IOS, Amsterdam).

[8] Asoka-Kumar P, Alatalo M, Gosh V J, Kruseman A C, Nielsen B and Lynn K G 1996 Phys. Rev. Lett. **77** 2097.

[9] Gosh V J, Alatalo M, Asoka-Kumar P, Nielsen B, Lynn K G, Kruseman A C and Mijnarends P E 2000 Phys. Rev. B **61** 10092.

[10] Rodda J L and Stewart M G 1963 Phys. Rev. **131** 255.

[11] MacKenzie I K, Jackman T E and Thrane N 1975 Phys. Rev. Lett. **34** 512.

[12] Brandt W, Isaacson D and Paulin R 1975 Phys. Rev. Lett. **35** 1180.

[13] Welch D O and Lynn K G 1976 Phys. Stat. Sol. (b) **77** 277.

[14] Puska M J 1991 J. Phys.: Condens. Matter **3** 3455.

[15] Myler U and Simpson P J 1997 Phys. Rev. B **56** 14303.

[16] Puska M J, Lanki P and Nieminen R M 1989 J. Phys.: Condens. Matter **1** 6081.

[17] Carbotte J P 1966 Phys. Rev. **144** 309.

[18] Carbotte J P and Salvadori A 1967 Phys. Rev. **162** 290.

[19] Hodges C H 1970 Phys. Rev. Lett. **25** 284.

[20] Hodges C H and Stott M J 1973 Phys. Rev. B **7** 73.

[21] Makkonen I, Hakala M and Puska M J 2006 Phys. Rev. B **73** 35103.

[22] Nieminen R M, Boronski E and Lantto L 1985 Phys. Rev. B **32** 1377.

[23] Boroński E and Nieminen R M 1986 Phys. Rev. B **34** 3820.

[24] Puska M J and Nieminen R M 1983 J. Phys. F: Met. Phys. **13** 333.

[25] Andersen O K, Jepsen O and Glötzel D 1985 *Highlight of Condensed-Matter Theory*, ed F Bassani, F Fumi and M P Tosi (North-Holland, New York).

[26] Andersen O K, Jepsen O and Šob M 1987 *Electronic Band Structure and Its Applications*, ed M Yossouff, (Springer-Verlag, Heildeberg) 1.

[27] Jensen K O 1989 J. Phys.: Condens. Matter **1** 10595.

[28] Barbiellini B, Puska M J, Korhonen T, Harju A, Torsti T and Nieminen R M 1996 Phys. Rev. B **53** 16201.

[29] Korhonen T, Puska M J and Nieminen R M 1996 Phys. Rev. B **54** 15016.

[30] Kimball G E and Shortley G H 1934 Phys. Rev. **45** 815.

[31] Arponen J and Pajanne E 1979 Ann. Phys. (N.Y.) **121** 343; Arponen J and Pajanne E 1985 *Positron Annihilation 7*, ed P C Jain, R M Singru and K P Gopinathan (World-Scientific, Singapore) 21.

[32] Lantto L 1987 Phys. Rev. B **36** 5160.

[33] Barbiellini B, Puska M J, Torsti T and Nieminen R M 1995 Phys. Rev. B **51** 7341.

[34] Campillo J M, Plazaola F and Puska M J 1998 Phys. Stat. Sol. (b) **206** 509.

[35] Laasonen K, Nieminen R and Puska M J 1992 Phys. Rev. B **45** 4122.

[36] Saito M and Oshiyama A 1996 Phys. Rev. B **53** 7810.

[37] Campillo J M, Plazaola F and de Diego N 2000 J. Phys.: Condens. Matter **12** 9715.

[38] Khanna S N, Rao B K, Jena P, Esterling D and Puska M J 1988 Phys. Rev. B **37** 6.

[39] Kamimura Y, Hori F, Tsutsumi T and Kuramoto E 1995 *Positron Annihilation 10*, ed Y-J He, B-S Cao and Y C Jean, Mater. Sci. Forum **175-178** 403.

[40] Donohue J 1974 *The Structures of the elements* (John Wiley & Sons).

[41] Kittel C 1996 *Introduction to Solid State Physics* (John Wiley & Sons).

[42] Ashcroft N W and Mermin N D 1976 *Solid State Physics* (Saunders College).

[43] Puska M J, Mäkinen S, Manninen M and Nieminen R M 1989 Phys. Rev. B **39** 7666.

[44] Landolt H and Börstein R 1985 *Numerical data and functional relationships in science and technology* (Springer Verlag, Berlin).

[45] Campillo Robles J M and Plazaola F 2003 Deffect and Diffusion Forum Vols. **213-215** 141.

[46] Shearer J W and Deutsch M 1949 Phys. Rev. **76** 462.

[47] Deutsch M 1951 Phys. Rev. **82** 455.

[48] Deutsch M 1951 Phys. Rev. **83** 207.

[49] Deutsch M 1951 Phys. Rev. **83** 866.

[50] Kirkegaard P and Eldrup M 1972 Computer Phys. Commun. **3** 240.

[51] Kirkegaard P and Eldrup M 1974 Computer Phys. Commun. **7** 401.

[52] Weisberg H and Berko S 1967 Phys. Rev. **154** 249.

[53] Liu D C and Roberts W K 1963 Phys. Rev. **132** 1633.

[54] Singman C N 1984 Jour. Chem. Educ. **61** n2 137.

[55] Pettifor D 1995 *Bonding and Structure of molecules and solids* (Oxford Univ. Press).

[56] Greenwood N N and Earnshaw A 1984 *Chemestry of the Elements* (Pergamon Press, Oxford).

[57] Porterfield W W 1993 *Inorganic Chemistry. A Unified Approach* (Academic Press, Inc.).

[58] Shriver D F, Atkins P W and Langford C H 1990 *Inorganic Chemestry* (Oxford Univ. Press).

[59] Rich R L 1972 *Periodic Correlations* (Addison-Wesley).

[60] Meyer J L 1870 Justus Liebigs Annalen der Chemie, supp. **7** 354.

[61] Rubaszek A, Szotek Z and Temmerman W M 1998 Phys. Rev. B **58** 11285.

[62] Stachowiak H, Boroński E and Banach G 2001 Acta Physica Polonica A **99** 503.

[63] Mitroy J and Barbiellini B 2002 Phys. Rev. B **65** 235103.

[64] Stachowiak H and Boroński E 2005 Acta Physica Polonica A **107** 541.

[65] Fraser L M September 1995 *Coulomb Interactions and Positron Annihilation in Many Fermion Systems: A Monte Carlo Approach*, Thesis, London (www.imperial.ac.uk/research/cmth/research/theses/L.M.Fraser.pdf ).

[66] Boroński E 2005 Acta Physica Polonica A **107** 576.

[67] Arponen J and Pajanne E 1979 J. Phys. F **9** 2359.

[68] Gondzik J and Stachowiak H 1985 J. Phys. C **18** 5399.

[69] Rubaszek A and Stachowiak H 1988 Phys. Rev. B **38** 3846.

[70] Barbiellini B 2001 *New Directions in Antimatter Chemestry and Physics*, ed C M Surko and F A Gianturco (Kluwer Academic Publishers) p 127.

[71] Stachowiak H and Lach J 1993 Phys. Rev. B **48** 9828.

[72] Ishibashi S and Kohyama M 2003 Phys. Rev. B **67** 113403.

[73] Skriver H 1983 *Systematics and Properties of the Lanthanides* ed S P Sinha (Reidel, Dordrecht) p 213.

[74] Johansson B 1979 Phys. Rev. B **20** 1315.

[75] Strange P, Svane A, Temmerman W M, Szotek Z and Winter H 1999 Nature **399** 756.

[76] Johansson B and Skriver H L 1982 J. Magn. Magn. Mat. **29** 217.

[77] Petit L, Svane A, Szotek Z and Temmerman W M 2003 Mol. Phys. Rep. **38** 20.

[78] Biasini M, Rusz J, Ferro G and Czopnik A 2005 Acta Physica Polonica A **107** 554.

[79] Plazaola F, Seitsonen A P and Puska M J 1994 J. Phys.: Condens. Matter **6** 8809.

## TABLE CAPTIONS

**Table 1.** Structural data of the elements. The crystal structures of the elements are named as: cubic (CUB), body-centered cubic (BCC), face-centered cubic (FCC), cubic diamond (DIAM), rhombohedral (RHOMB), tetragonal (TETRA), orthorhombic (ORTHOR), hexagonal close-packed (HEX) and double hexagonal close-packed (HEX/abac). (*) Manganese has a cubic complex structure (see Donohue [40]).

**Table 2.** High-frequency dielectric constant of some elements used in calculations.

**Table 3.** Calculated positron bulk lifetimes using AT-SUP method with BN and GGA parametrizations.

**Table 4.** Calculated positron bulk lifetimes using LMTO method with BN and GGA parametrizations.

**Table 5.** Calculated positron monovacancy lifetimes using AT-SUP method with BN and GGA parametrizations.

**Table 6.** Calculated positron monovacancy lifetimes using LMTO method with BN and GGA parametrizations.

**Table 7.** Calculated positron lifetimes using AT-SUP method with LDA parametrization for bulk and monovacancy states.

**Table 8.** Experimental positron lifetimes for bulk and monovacancy states.

## TABLE 1

Element (U); Crystal Structure (ORTHOR); *a* (Å) 2.85; *b* (Å) 5.87; *c* (Å) 4.96

| | | | | | | | | | | | | | | | | | |
|---|---|---|---|---|---|---|---|---|---|---|---|---|---|---|---|---|---|
| **H**<br>—<br>—<br>—<br>— | | | | | | | | | | | | | | | | | **He**<br>—<br>—<br>—<br>— |
| **Li**<br>BCC<br>3.49 | **Be**<br>HEX<br>2.29<br>3.59 | | | | | | | | | | | **B**<br>—<br>—<br>—<br>— | **C**<br>DIAM<br>3.57 | **N**<br>—<br>—<br>—<br>— | **O**<br>—<br>—<br>—<br>— | **F**<br>—<br>—<br>—<br>— | **Ne**<br>FCC<br>4.43 |
| **Na**<br>BCC<br>4.23 | **Mg**<br>HEX<br>3.21<br>5.21 | | | | | | | | | | | **Al**<br>FCC<br>4.05 | **Si**<br>DIAM<br>5.43 | **P**<br>ORTHOR<br>3.32<br>10.52<br>4.39 | **S**<br>—<br>—<br>—<br>— | **Cl**<br>ORTHOR<br>6.24<br>4.48<br>8.26 | **Ar**<br>FCC<br>5.31 |
| **K**<br>BCC<br>5.23 | **Ca**<br>FCC<br>5.58 | **Sc**<br>HEX<br>3.31<br>5.28 | **Ti**<br>HEX<br>2.95<br>4.68 | **V**<br>BCC<br>3.03 | **Cr**<br>BCC<br>2.88 | **Mn**<br>CUB (*)<br>8.91 | **α-Fe**<br>BCC<br>2.87 | **α-Co**<br>HEX<br>2.51<br>4.07 | **Ni**<br>FCC<br>3.52 | **Cu**<br>FCC<br>3.61 | **Zn**<br>HEX<br>2.66<br>4.94 | **Ga**<br>ORTHOR<br>4.52<br>7.66<br>4.53 | **Ge**<br>DIAM<br>5.66 | **As**<br>RHOMB<br>3.76<br>10.55 | **Se**<br>HEX<br>4.36<br>4.95 | **Br**<br>ORTHOR<br>6.67<br>4.48<br>8.72 | **Kr**<br>FCC<br>5.64 |
| **Rb**<br>BCC<br>5.59 | **Sr**<br>FCC<br>6.08 | **Y**<br>HEX<br>3.65<br>5.73 | **Zr**<br>HEX<br>3.23<br>5.15 | **Nb**<br>BCC<br>3.30 | **Mo**<br>BCC<br>3.15 | **Tc**<br>HEX<br>2.74<br>4.40 | **Ru**<br>HEX<br>2.70<br>4.28 | **Rh**<br>FCC<br>3.80 | **Pd**<br>FCC<br>3.89 | **Ag**<br>FCC<br>4.09 | **Cd**<br>HEX<br>2.98<br>5.62 | **In**<br>TETRA<br>3.25<br>4.95 | **β-Sn**<br>TETRA<br>5.83<br>3.18 | **Sb**<br>RHOMB<br>4.31<br>11.27 | **Te**<br>HEX<br>4.45<br>5.92 | **I**<br>ORTHOR<br>7.27<br>4.80<br>9.80 | **Xe**<br>FCC<br>6.13 |
| **Cs**<br>BCC<br>6.05 | **Ba**<br>BCC<br>5.02 | **La**<br>HEX<br>3.75<br>6.07 | **Hf**<br>HEX<br>3.20<br>5.06 | **Ta**<br>BCC<br>3.31 | **W**<br>BCC<br>3.16 | **Re**<br>HEX<br>2.76<br>4.46 | **Os**<br>HEX<br>2.74<br>4.33 | **Ir**<br>FCC<br>3.84 | **Pt**<br>FCC<br>3.92 | **Au**<br>FCC<br>4.08 | **β-Hg**<br>TETRA<br>4.00<br>2.83 | **Tl**<br>HEX<br>3.46<br>5.53 | **Pb**<br>FCC<br>4.95 | **Bi**<br>RHOMB<br>4.55<br>11.86 | **Po**<br>CUB<br>3.35 | **At**<br>—<br>—<br>—<br>— | **Rn**<br>—<br>—<br>—<br>— |
| **Fr**<br>—<br>—<br>—<br>— | **Ra**<br>BCC<br>5.15 | **Ac**<br>FCC<br>5.31 | | | | | | | | | | | | | | | |

| | | | | | | | | | | | | | |
|---|---|---|---|---|---|---|---|---|---|---|---|---|---|
| **γ-Ce**<br>FCC<br>5.16 | **Pr**<br>HEX/abac<br>3.67<br>11.83 | **Nd**<br>HEX<br>3.66<br>5.91 | **Pm**<br>HEX/abac<br>3.65<br>11.65 | **Sm**<br>RHOMB<br>3.63<br>26.22 | **Eu**<br>BCC<br>4.61 | **Gd**<br>HEX<br>3.64<br>5.78 | **Tb**<br>HEX<br>3.60<br>5.69 | **Dy**<br>HEX<br>3.59<br>5.65 | **Ho**<br>HEX<br>3.58<br>5.62 | **Er**<br>HEX<br>3.56<br>5.59 | **Tm**<br>HEX<br>3.54<br>5.56 | **Yb**<br>FCC<br>5.49 | **Lu**<br>HEX<br>3.51<br>5.56 |
| **Th**<br>FCC<br>5.08 | **Pa**<br>TETRA<br>3.93<br>3.24 | **U**<br>ORTHOR<br>2.85<br>5.87<br>4.96 | **Np**<br>ORTHOR<br>4.72<br>4.89<br>6.66 | **Pu**<br>FCC<br>4.64 | **Am**<br>HEX/abac<br>3.47<br>11.24 | **Cm**<br>HEX/abac<br>3.50<br>11.33 | **Bk**<br>FCC<br>5.00 | **Cf**<br>—<br>—<br>—<br>— | **Es**<br>—<br>—<br>—<br>— | **Fm**<br>—<br>—<br>—<br>— | **Md**<br>—<br>—<br>—<br>— | **No**<br>—<br>—<br>—<br>— | **Lr**<br>—<br>—<br>—<br>— |

**TABLE 2**

| **Element** | $\varepsilon_\infty$ |
|---|---|
| C | 5.62 |
| Si | 11.9 |
| P | 6.1 |
| Ge | 16.0 |
| Se | 13.98 |
| Sn | 23.8 |
| Te | 29.5 |

**TABLE 3**

**Element** → **Zn**: 139 = $\tau_{BN}^{AT-SUP}(ps)$; 158 = $\tau_{GGA}^{AT-SUP}(ps)$ — **Bulk Lifetimes**

| | | | | | | | | | | | | | | | | | |
|---|---|---|---|---|---|---|---|---|---|---|---|---|---|---|---|---|---|
| **H**<br>—<br>— | | | | | | | | | | | | | | | | | **He**<br>—<br>— |
| **Li**<br>301<br>285 | **Be**<br>137<br>129 | | | | | | | | | | | **B**<br>—<br>— | **C**<br>93<br>93 | **N**<br>—<br>— | **O**<br>—<br>— | **F**<br>—<br>— | **Ne**<br>237<br>566 |
| **Na**<br>322<br>342 | **Mg**<br>233<br>226 | | | | | | | | | | | **Al**<br>168<br>160 | **Si**<br>218<br>207 | **P**<br>230<br>226 | **S**<br>—<br>— | **Cl**<br>248<br>352 | **Ar**<br>271<br>483 |
| **K**<br>367<br>402 | **Ca**<br>288<br>281 | **Sc**<br>197<br>199 | **Ti**<br>147<br>153 | **V**<br>116<br>124 | **Cr**<br>104<br>118 | **Mn**<br>106<br>115 | **α-Fe**<br>102<br>112 | **α-Co**<br>97<br>108 | **Ni**<br>96<br>108 | **Cu**<br>108<br>130 | **Zn**<br>139<br>158 | **Ga**<br>190<br>202 | **Ge**<br>222<br>228 | **As**<br>184<br>195 | **Se**<br>286<br>355 | **Br**<br>245<br>322 | **Kr**<br>281<br>459 |
| **Rb**<br>376<br>420 | **Sr**<br>309<br>305 | **Y**<br>217<br>215 | **Zr**<br>160<br>162 | **Nb**<br>126<br>134 | **Mo**<br>109<br>118 | **Tc**<br>98<br>105 | **Ru**<br>94<br>106 | **Rh**<br>97<br>110 | **Pd**<br>107<br>130 | **Ag**<br>125<br>150 | **Cd**<br>159<br>184 | **In**<br>183<br>201 | **β-Sn**<br>193<br>201 | **Sb**<br>213<br>227 | **Te**<br>289<br>345 | **I**<br>262<br>325 | **Xe**<br>297<br>444 |
| **Cs**<br>387<br>437 | **Ba**<br>305<br>303 | **La**<br>209<br>209 | **Hf**<br>150<br>156 | **Ta**<br>119<br>125 | **W**<br>102<br>108 | **Re**<br>94<br>101 | **Os**<br>89<br>96 | **Ir**<br>90<br>98 | **Pt**<br>99<br>116 | **Au**<br>112<br>131 | **β-Hg**<br>154<br>184 | **Tl**<br>185<br>214 | **Pb**<br>190<br>214 | **Bi**<br>230<br>257 | **Po**<br>231<br>262 | **At**<br>—<br>— | **Rn**<br>—<br>— |
| **Fr**<br>—<br>— | **Ra**<br>307<br>308 | **Ac**<br>199<br>197 | | | | | | | | | | | | | | | |
| | | | | **γ-Ce**<br>204<br>210 | **Pr**<br>207<br>213 | **Nd**<br>207<br>214 | **Pm**<br>206<br>214 | **Sm**<br>206<br>214 | **Eu**<br>273<br>278 | **Gd**<br>204<br>207 | **Tb**<br>204<br>214 | **Dy**<br>204<br>214 | **Ho**<br>204<br>215 | **Er**<br>202<br>214 | **Tm**<br>201<br>213 | **Yb**<br>256<br>265 | **Lu**<br>196<br>202 |
| | | | | **Th**<br>173<br>171 | **Pa**<br>139<br>142 | **U**<br>117<br>121 | **Np**<br>119<br>125 | **Pu**<br>145<br>155 | **Am**<br>171<br>182 | **Cm**<br>172<br>180 | **Bk**<br>176<br>181 | **Cf**<br>—<br>— | **Es**<br>—<br>— | **Fm**<br>—<br>— | **Md**<br>—<br>— | **No**<br>—<br>— | **Lr**<br>—<br>— |

**TABLE 4**

**Element** → **Zn**: 134 ← **Bulk Lifetimes** $\tau_{BN}^{LMTO}(ps)$; 146 ← $\tau_{GGA}^{LMTO}(ps)$

| | | | | | | | | | | | | | | | | | |
|---|---|---|---|---|---|---|---|---|---|---|---|---|---|---|---|---|---|
| **H**<br>—<br>— | | | | | | | | | | | | | | | | | **He**<br>—<br>— |
| **Li**<br>301<br>281 | **Be**<br>137<br>128 | | | | | | | | | | | **B**<br>—<br>— | **C**<br>99<br>97 | **N**<br>—<br>— | **O**<br>—<br>— | **F**<br>—<br>— | **Ne**<br>206<br>534 |
| **Na**<br>324<br>334 | **Mg**<br>232<br>216 | | | | | | | | | | | **Al**<br>165<br>153 | **Si**<br>222<br>211 | **P**<br>237<br>223 | **S**<br>—<br>— | **Cl**<br>250<br>275 | **Ar**<br>268<br>435 |
| **K**<br>368<br>391 | **Ca**<br>290<br>279 | **Sc**<br>196<br>193 | **Ti**<br>144<br>146 | **V**<br>114<br>119 | **Cr**<br>99<br>105 | **Mn**<br>102<br>109 | **α-Fe**<br>101<br>110 | **α-Co**<br>96<br>107 | **Ni**<br>96<br>108 | **Cu**<br>105<br>118 | **Zn**<br>134<br>146 | **Ga**<br>165<br>167 | **Ge**<br>223<br>228 | **As**<br>181<br>188 | **Se**<br>281<br>313 | **Br**<br>248<br>270 | **Kr**<br>289<br>448 |
| **Rb**<br>377<br>408 | **Sr**<br>311<br>300 | **Y**<br>215<br>208 | **Zr**<br>156<br>154 | **Nb**<br>120<br>122 | **Mo**<br>103<br>106 | **Tc**<br>94<br>99 | **Ru**<br>89<br>95 | **Rh**<br>92<br>100 | **Pd**<br>102<br>113 | **Ag**<br>122<br>137 | **Cd**<br>154<br>167 | **In**<br>181<br>190 | **β-Sn**<br>182<br>179 | **Sb**<br>208<br>211 | **Te**<br>286<br>307 | **I**<br>260<br>268 | **Xe**<br>299<br>416 |
| **Cs**<br>387<br>426 | **Ba**<br>307<br>304 | **La**<br>208<br>207 | **Hf**<br>146<br>146 | **Ta**<br>115<br>117 | **W**<br>98<br>100 | **Re**<br>90<br>93 | **Os**<br>85<br>89 | **Ir**<br>86<br>92 | **Pt**<br>93<br>101 | **Au**<br>108<br>119 | **β-Hg**<br>146<br>161 | **Tl**<br>183<br>203 | **Pb**<br>188<br>197 | **Bi**<br>224<br>233 | **Po**<br>214<br>216 | **At**<br>—<br>— | **Rn**<br>—<br>— |
| **Fr**<br>—<br>— | **Ra**<br>308<br>309 | **Ac**<br>195<br>192 | | | | | | | | | | | | | | | |
| | | | | **γ-Ce**<br>197<br>196 | **Pr**<br>200<br>200 | **Nd**<br>202<br>202 | **Pm**<br>201<br>202 | **Sm**<br>200<br>203 | **Eu**<br>271<br>266 | **Gd**<br>205<br>209 | **Tb**<br>201<br>205 | **Dy**<br>200<br>205 | **Ho**<br>200<br>206 | **Er**<br>199<br>205 | **Tm**<br>198<br>203 | **Yb**<br>254<br>253 | **Lu**<br>193<br>192 |
| | | | | **Th**<br>170<br>167 | **Pa**<br>125<br>125 | **U**<br>107<br>108 | **Np**<br>116<br>117 | **Pu**<br>137<br>141 | **Am**<br>164<br>171 | **Cm**<br>171<br>180 | **Bk**<br>181<br>193 | **Cf**<br>—<br>— | **Es**<br>—<br>— | **Fm**<br>—<br>— | **Md**<br>—<br>— | **No**<br>—<br>— | **Lr**<br>—<br>— |

## TABLE 5

Element: Monovacancy Lifetimes $\tau_{BN}^{AT-SUP}(ps)$, $\tau_{GGA}^{AT-SUP}(ps)$ (**Zn** 196 / 224)

| | | | | | | | | | | | | | | | | | |
|---|---|---|---|---|---|---|---|---|---|---|---|---|---|---|---|---|---|
| **H**<br>—<br>— | | | | | | | | | | | | | | | | | **He**<br>—<br>— |
| **Li**<br>338<br>315 | **Be**<br>180<br>165 | | | | | | | | | | | **B**<br>—<br>— | **C**<br>110<br>109 | **N**<br>—<br>— | **O**<br>—<br>— | **F**<br>—<br>— | **Ne**<br>239<br>562 |
| **Na**<br>362<br>384 | **Mg**<br>299<br>292 | | | | | | | | | | | **Al**<br>244<br>231 | **Si**<br>252<br>241 | **P**<br>251<br>245 | **S**<br>—<br>— | **Cl**<br>249<br>353 | **Ar**<br>274<br>483 |
| **K**<br>411<br>452 | **Ca**<br>366<br>367 | **Sc**<br>291<br>281 | **Ti**<br>236<br>228 | **V**<br>198<br>194 | **Cr**<br>180<br>188 | **Mn**<br>198<br>199 | **α-Fe**<br>177<br>183 | **α-Co**<br>168<br>177 | **Ni**<br>166<br>177 | **Cu**<br>169<br>200 | **Zn**<br>196<br>224 | **Ga**<br>239<br>255 | **Ge**<br>254<br>263 | **As**<br>236<br>255 | **Se**<br>297<br>359 | **Br**<br>251<br>324 | **Kr**<br>287<br>458 |
| **Rb**<br>419<br>475 | **Sr**<br>387<br>398 | **Y**<br>319<br>311 | **Zr**<br>262<br>254 | **Nb**<br>222<br>224 | **Mo**<br>200<br>206 | **Tc**<br>185<br>186 | **Ru**<br>177<br>191 | **Rh**<br>178<br>198 | **Pd**<br>173<br>220 | **Ag**<br>200<br>245 | **Cd**<br>231<br>276 | **In**<br>273<br>303 | **β-Sn**<br>277<br>294 | **Sb**<br>275<br>303 | **Te**<br>329<br>353 | **I**<br>290<br>329 | **Xe**<br>320<br>444 |
| **Cs**<br>429<br>496 | **Ba**<br>392<br>399 | **La**<br>322<br>310 | **Hf**<br>254<br>252 | **Ta**<br>217<br>217 | **W**<br>194<br>197 | **Re**<br>183<br>188 | **Os**<br>175<br>184 | **Ir**<br>175<br>189 | **Pt**<br>177<br>191 | **Au**<br>191<br>233 | **β-Hg**<br>218<br>270 | **Tl**<br>268<br>325 | **Pb**<br>277<br>329 | **Bi**<br>288<br>335 | **Po**<br>293<br>345 | **At**<br>—<br>— | **Rn**<br>—<br>— |
| **Fr**<br>—<br>— | **Ra**<br>394<br>415 | **Ac**<br>317<br>306 | | | | | | | | | | | | | | | |
| | | | | | | | | | | | | | | | | | |
| | | | | **γ-Ce**<br>316<br>302 | **Pr**<br>318<br>306 | **Nd**<br>318<br>306 | **Pm**<br>316<br>307 | **Sm**<br>314<br>307 | **Eu**<br>366<br>378 | **Gd**<br>312<br>309 | **Tb**<br>311<br>307 | **Dy**<br>310<br>307 | **Ho**<br>309<br>307 | **Er**<br>307<br>307 | **Tm**<br>305<br>306 | **Yb**<br>346<br>364 | **Lu**<br>300<br>301 |
| | | | | **Th**<br>292<br>284 | **Pa**<br>250<br>238 | **U**<br>218<br>208 | **Np**<br>203<br>199 | **Pu**<br>259<br>255 | **Am**<br>285<br>287 | **Cm**<br>287<br>289 | **Bk**<br>290<br>299 | **Cf**<br>—<br>— | **Es**<br>—<br>— | **Fm**<br>—<br>— | **Md**<br>—<br>— | **No**<br>—<br>— | **Lr**<br>—<br>— |

**TABLE 6**

**Element** → **Zn** 217 / 229 ← **Monovacancy Lifetimes** $\tau_{BN}^{LMTO}(ps)$ / $\tau_{GGA}^{LMTO}(ps)$

| | | | | | | | | | | | | | | | | | |
|---|---|---|---|---|---|---|---|---|---|---|---|---|---|---|---|---|---|
| **H**<br>—<br>— | | | | | | | | | | | | | | | | | **He**<br>—<br>— |
| **Li**<br>330<br>308 | **Be**<br>183<br>161 | | | | | | | | | | | **B**<br>—<br>— | **C**<br>118<br>114 | **N**<br>—<br>— | **O**<br>—<br>— | **F**<br>—<br>— | **Ne**<br>244<br>556 |
| **Na**<br>363<br>377 | **Mg**<br>308<br>292 | | | | | | | | | | | **Al**<br>250<br>237 | **Si**<br>262<br>245 | **P**<br>275<br>260 | **S**<br>—<br>— | **Cl**<br>280<br>321 | **Ar**<br>290<br>456 |
| **K**<br>409<br>446 | **Ca**<br>374<br>364 | **Sc**<br>298<br>282 | **Ti**<br>242<br>234 | **V**<br>203<br>200 | **Cr**<br>184<br>184 | **Mn**<br>194<br>196 | **α-Fe**<br>181<br>188 | **α-Co**<br>172<br>182 | **Ni**<br>169<br>182 | **Cu**<br>178<br>194 | **Zn**<br>217<br>229 | **Ga**<br>238<br>240 | **Ge**<br>265<br>266 | **As**<br>253<br>267 | **Se**<br>326<br>392 | **Br**<br>300<br>337 | **Kr**<br>323<br>482 |
| **Rb**<br>415<br>466 | **Sr**<br>388<br>393 | **Y**<br>324<br>311 | **Zr**<br>269<br>258 | **Nb**<br>226<br>220 | **Mo**<br>205<br>204 | **Tc**<br>191<br>191 | **Ru**<br>183<br>187 | **Rh**<br>185<br>194 | **Pd**<br>192<br>209 | **Ag**<br>212<br>236 | **Cd**<br>251<br>273 | **In**<br>278<br>298 | **β-Sn**<br>280<br>283 | **Sb**<br>290<br>307 | **Te**<br>350<br>400 | **I**<br>329<br>370 | **Xe**<br>353<br>473 |
| **Cs**<br>423<br>499 | **Ba**<br>395<br>395 | **La**<br>328<br>321 | **Hf**<br>261<br>253 | **Ta**<br>224<br>219 | **W**<br>203<br>203 | **Re**<br>191<br>191 | **Os**<br>184<br>188 | **Ir**<br>185<br>193 | **Pt**<br>192<br>206 | **Au**<br>206<br>228 | **β-Hg**<br>234<br>264 | **Tl**<br>278<br>320 | **Pb**<br>293<br>324 | **Bi**<br>299<br>318 | **Po**<br>312<br>346 | **At**<br>—<br>— | **Rn**<br>—<br>— |
| **Fr**<br>—<br>— | **Ra**<br>393<br>426 | **Ac**<br>324<br>324 | | | | | | | | | | | | | | | |
| | | | | **γ-Ce**<br>315<br>312 | **Pr**<br>317<br>312 | **Nd**<br>317<br>312 | **Pm**<br>315<br>311 | **Sm**<br>314<br>311 | **Eu**<br>362<br>373 | **Gd**<br>316<br>314 | **Tb**<br>312<br>308 | **Dy**<br>311<br>308 | **Ho**<br>311<br>307 | **Er**<br>309<br>305 | **Tm**<br>307<br>304 | **Yb**<br>350<br>355 | **Lu**<br>305<br>297 |

**TABLE 7**

**Element**

| | |
|---|---|
| **Ga**<br>168<br>210 | $\tau_{bulk}^{LDA}(ps)$<br>$\tau_{vac}^{LDA}(ps)$ |

| | | | | | | | | | | | | | | | | | |
|---|---|---|---|---|---|---|---|---|---|---|---|---|---|---|---|---|---|
| **H**<br>—<br>— | | | | | | | | | | | | | | | | | **He**<br>—<br>— |
| **Li**<br>261<br>294 | **Be**<br>123<br>158 | | | | | | | | | | | **B**<br>—<br>— | **C**<br>84<br>98 | **N**<br>—<br>— | **O**<br>—<br>— | **F**<br>—<br>— | **Ne**<br>222<br>223 |
| **Na**<br>286<br>326 | **Mg**<br>203<br>260 | | | | | | | | | | | **Al**<br>148<br>212 | **Si**<br>184<br>210 | **P**<br>188<br>205 | **S**<br>—<br>— | **Cl**<br>221<br>223 | **Ar**<br>246<br>250 |
| **K**<br>332<br>387 | **Ca**<br>249<br>326 | **Sc**<br>173<br>253 | **Ti**<br>132<br>206 | **V**<br>106<br>174 | **Cr**<br>96<br>160 | **Mn**<br>98<br>175 | **α-Fe**<br>94<br>157 | **α-Co**<br>90<br>150 | **Ni**<br>90<br>149 | **Cu**<br>100<br>153 | **Zn**<br>125<br>177 | **Ga**<br>168<br>210 | **Ge**<br>190<br>216 | **As**<br>163<br>208 | **Se**<br>244<br>257 | **Br**<br>218<br>223 | **Kr**<br>253<br>260 |
| **Rb**<br>343<br>401 | **Sr**<br>269<br>350 | **Y**<br>189<br>278 | **Zr**<br>142<br>228 | **Nb**<br>114<br>195 | **Mo**<br>100<br>176 | **Tc**<br>91<br>163 | **Ru**<br>88<br>158 | **Rh**<br>90<br>159 | **Pd**<br>99<br>157 | **Ag**<br>115<br>180 | **Cd**<br>143<br>205 | **In**<br>162<br>241 | **β-Sn**<br>171<br>243 | **Sb**<br>187<br>242 | **Te**<br>250<br>289 | **I**<br>231<br>260 | **Xe**<br>267<br>294 |
| **Cs**<br>356<br>420 | **Ba**<br>265<br>355 | **La**<br>183<br>281 | **Hf**<br>135<br>221 | **Ta**<br>109<br>191 | **W**<br>94<br>172 | **Re**<br>88<br>162 | **Os**<br>83<br>156 | **Ir**<br>84<br>157 | **Pt**<br>92<br>160 | **Au**<br>103<br>172 | **β-Hg**<br>139<br>196 | **Tl**<br>165<br>239 | **Pb**<br>169<br>247 | **Bi**<br>202<br>256 | **Po**<br>204<br>261 | **At**<br>—<br>— | **Rn**<br>—<br>— |
| **Fr**<br>—<br>— | **Ra**<br>267<br>360 | **Ac**<br>174<br>276 | | | | | | | | | | | | | | | |
| | | | | **Ce**<br>179<br>275 | **Pr**<br>182<br>277 | **Nd**<br>182<br>277 | **Pm**<br>181<br>275 | **Sm**<br>181<br>274 | **Eu**<br>238<br>327 | **Gd**<br>179<br>273 | **Tb**<br>180<br>272 | **Dy**<br>179<br>271 | **Ho**<br>179<br>270 | **Er**<br>178<br>268 | **Tm**<br>177<br>267 | **Yb**<br>223<br>308 | **Lu**<br>172<br>262 |
| | | | | **Th**<br>153<br>254 | **Pa**<br>125<br>217 | **U**<br>107<br>190 | **Np**<br>109<br>178 | **Pu**<br>131<br>225 | **Am**<br>152<br>249 | **Cm**<br>153<br>251 | **Bk**<br>156<br>254 | **Cf**<br>—<br>— | **Es**<br>—<br>— | **Fm**<br>—<br>— | **Md**<br>—<br>— | **No**<br>—<br>— | **Lr**<br>—<br>— |

**TABLE 8**

**Element** → **Zn** 153 ← $\tau_{bulk}^{\exp}(ps)$; 220 ← $\tau_{vac}^{\exp}(ps)$

| | | | | | | | | | | | | | | | | | |
|---|---|---|---|---|---|---|---|---|---|---|---|---|---|---|---|---|---|
| **H**<br>—<br>— | | | | | | | | | | | | | | | | | **He**<br>—<br>— |
| **Li**<br>291<br>— | **Be**<br>137<br>— | | | | | | | | | | | **B**<br>—<br>— | **C**<br>107<br>— | **N**<br>—<br>— | **O**<br>—<br>— | **F**<br>—<br>— | **Ne**<br>—<br>— |
| **Na**<br>338<br>— | **Mg**<br>225<br>254 | | | | | | | | | | | **Al**<br>165<br>244 | **Si**<br>219<br>272 | **P** | **S**<br>—<br>— | **Cl** | **Ar**<br>430<br>— |
| **K**<br>397<br>— | **Ca** | **Sc**<br>230<br>— | **Ti**<br>150<br>222 | **V**<br>124<br>191 | **Cr**<br>120<br>150 | **Mn** | **α-Fe**<br>111<br>175 | **α-Co**<br>119<br>— | **Ni**<br>109<br>180 | **Cu**<br>120<br>180 | **Zn**<br>153<br>220 | **Ga**<br>198<br>— | **Ge**<br>228<br>279 | **As** | **Se**<br>335<br>— | **Br** | **Kr** |
| **Rb**<br>406<br>— | **Sr** | **Y**<br>249<br>— | **Zr**<br>164<br>252 | **Nb**<br>120<br>210 | **Mo**<br>106<br>170 | **Tc** | **Ru** | **Rh** | **Pd**<br>98<br>— | **Ag**<br>130<br>208 | **Cd**<br>184<br>252 | **In**<br>196<br>270 | **β-Sn**<br>200<br>242 | **Sb**<br>214<br>275 | **Te** | **I** | **Xe**<br>400<br>— |
| **Cs**<br>418<br>— | **Ba** | **La**<br>241<br>— | **Hf** | **Ta**<br>120<br>203 | **W**<br>105<br>195 | **Re** | **Os** | **Ir** | **Pt**<br>99<br>168 | **Au**<br>116<br>205 | **β-Hg** | **Tl**<br>226<br>258 | **Pb**<br>204<br>294 | **Bi**<br>240<br>325 | **Po** | **At**<br>—<br>— | **Rn**<br>—<br>— |
| **Fr**<br>—<br>— | **Ra** | **Ac** | | | | | | | | | | | | | | | |
| | | | | **γ-Ce** | **Pr** | **Nd** | **Pm** | **Sm**<br>199<br>— | **Eu** | **Gd**<br>230<br>— | **Tb** | **Dy** | **Ho** | **Er** | **Tm** | **Yb** | **Lu** |

## FIGURE CAPTIONS

**Figure 1.** Atomic volume (filled circles) and positron lifetimes plotted against atomic number. Positron lifetimes are calculated in bulk (circles) and monovacancy (squares) states within the AT-SUP method using BN approximation.

**Figure 2.** Bulk (circles) and monovacancy (squares) positron lifetimes of elements from La to Hg (6th row of the Periodic Table) versus atomic number. Lifetimes have been calculated within the LMTO using BN (open symbols) and GGA (filled symbols) approximations.

**FIGURE 1**

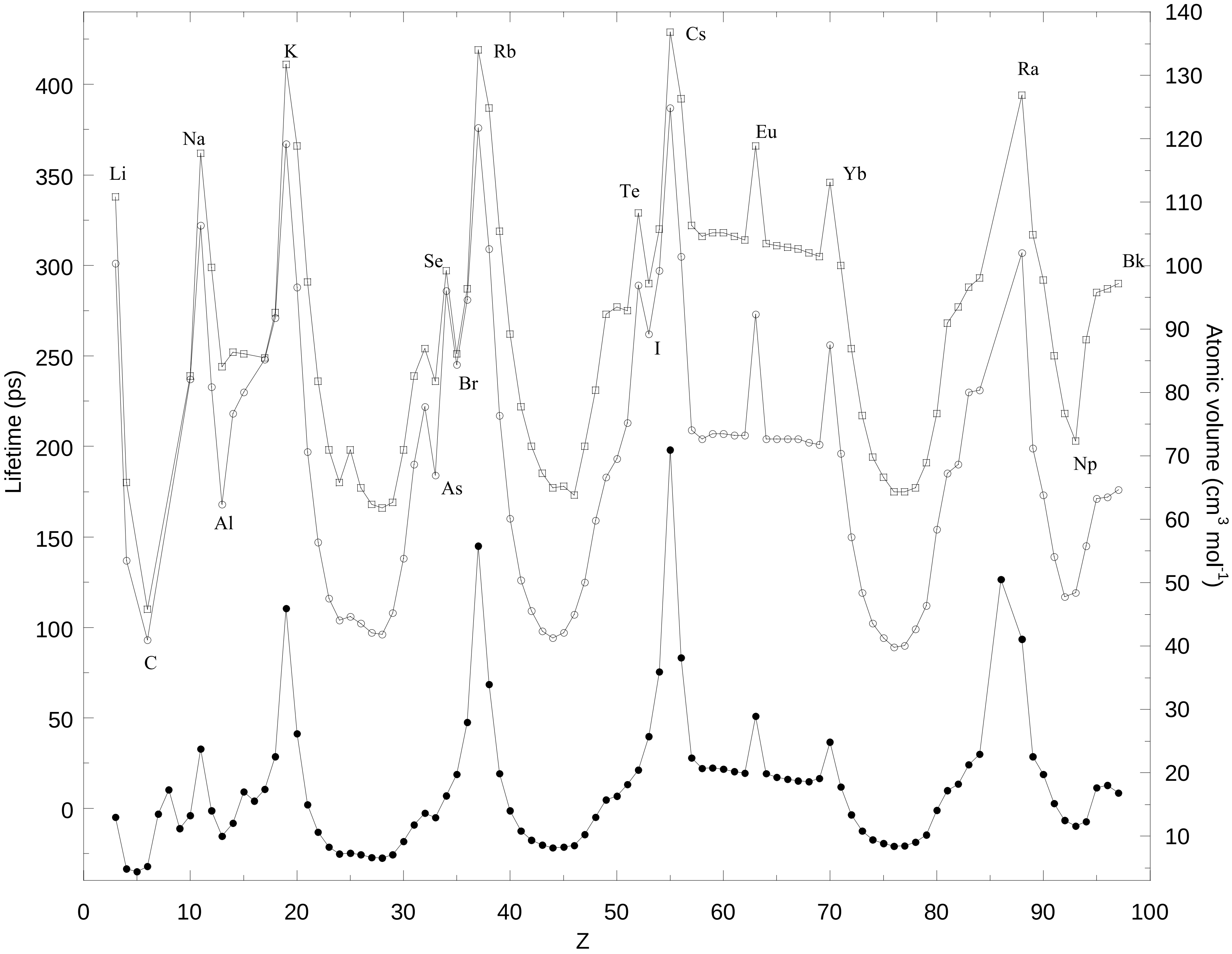

Li
Na
K
Rb
Cs
Eu
Yb
Ra
Bk
Te
I
Se
Br
As
Al
C
Np
Lifetime (ps)
Atomic volume (cm$^3$ mol$^{-1}$)
Z

**FIGURE 2**

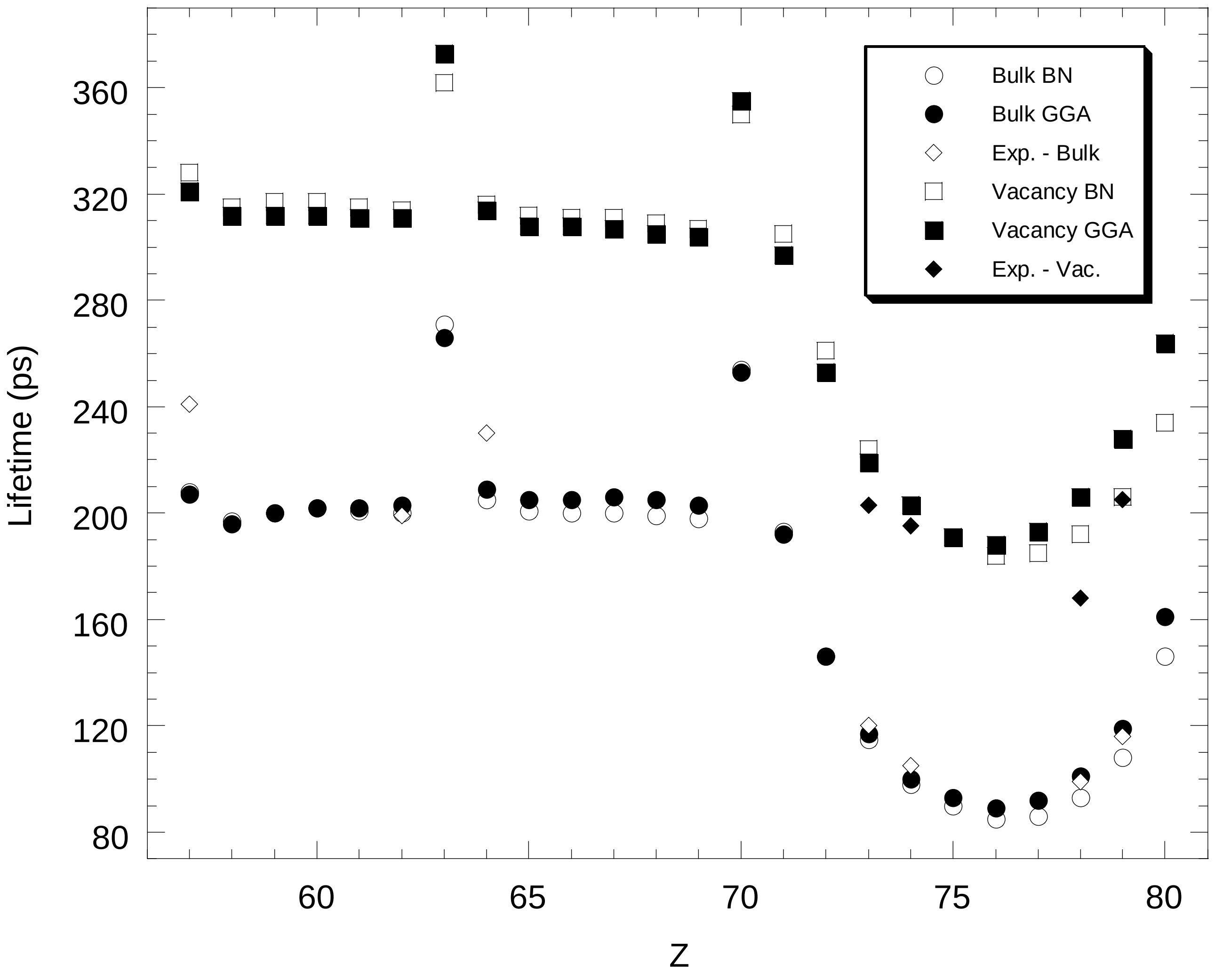

Bulk BN
Bulk GGA
Exp. - Bulk
Vacancy BN
Vacancy GGA
Exp. - Vac.
Lifetime (ps)
360
320
280
240
200
160
120
80
60
65
70
75
80
Z